\documentclass[notitlepage,12 pt,twoside,aps,prd,amsmath,amssymb, tightenlines,showpacs,showkeys,eqsecnum]{revtex4-1}
\pdfoutput=1
\usepackage{graphicx}
\usepackage{mathptmx}
\usepackage{bm}
\newcommand {\cG}{\cal G}
\newcommand {\cD}{\cal D}
\newcommand {\bg}{\bar \gamma}
\newcommand {\G}{\Gamma}
\newcommand {\bp}{\bar \psi}

\def\myfigure #1#2#3#4
{\begin{figure}[ht]\begin{center}
\includegraphics[width=#2 \textwidth]{#1.eps}
\parbox[t]{#4\textwidth}{\caption{#3}\label{#1}}
\end{center}\end{figure}}

\def \myfigures #1#2#3#4#5#6#7#8
{\begin{figure}[ht]
    \begin{center}
        \includegraphics[width=#2 \textwidth]{#1.jpg}
        \hfill
        \includegraphics[width=#6 \textwidth]{#5.jpg}
        \parbox[t]{#4\textwidth}{\vskip 10 mm \caption {#3}\label{#1}}
        \hfill
        \parbox[t]{#8\textwidth}{\vskip 10 mm \caption {#7}\label{#5}}
    \end{center}
\end{figure} }

\begin{document}

\baselineskip -24pt
\title{Spinor field in Bianchi type-$II$ Universe}
\author{Bijan Saha}
\affiliation{Laboratory of Information Technologies\\
Joint Institute for Nuclear Research\\
141980 Dubna, Moscow region, Russia} \email{bijan@jinr.ru}
\homepage{http://spinor.bijansaha.ru}

\hskip 1 cm

\begin{abstract}
Within the scope of Bianchi type-$II$ cosmological model we have
studied the role of spinor field in the evolution of the Universe.
It is found that in this case the components of energy-momentum
tensor of spinor field along the principal axis are different, i.e.
$T_1^1 \ne T_2^2 \ne T_3^3$, even in absence of spinor field
nonlinearity. The presence of nontrivial non-diagonal components of
energy-momentum tensor of the spinor field imposes severe
restrictions both on geometry of space-time and on the spinor field
itself. Depending on the choice of $f(x_3)$ the space-time might be
either locally rotationally symmetric or isotropic. In this paper we
considered the Bianchi type-$II$ space-time both for a trivial and
nontrivial $b$. It was found that while a positive $\lambda_1$ in
case of $f^\prime \ne 0$ gives rise to oscillatory mode of
evolution, in case of $f^\prime = 0$ it leads to accelerated mode of
expansion.
\end{abstract}

\keywords{Spinor field, dark energy, anisotropic cosmological
models, isotropization}

\pacs{98.80.Cq}

\maketitle

\bigskip

\section{Introduction}

In a number of papers it was shown that the introduction of
nonlinear spinor field as the source of gravity might be the answer
to some fundamental questions such as initial singularity,
isotropization, late time accelerated mode of expansion of the
Universe
\cite{Saha1997GRG,Saha1997JMP,Saha2001PRD,Saha2004aPRD,Saha2004bPRD,PopPLB,
PopPRD,PopGREG,FabIJTP,Saha2006ECAA,kremer1,Saha2006PRD,Saha2006GnC,Saha2007RRP,
Saha2009aECAA,ELKO,FabGRG}. Moreover, it was also shown that
different types of dark energy and perfect fluid can be simulated by
spinor field nonlinearity
\cite{Krechet,Saha2010CEJP,Saha2010RRP,Saha2011APSS,Saha2012IJTP}.
But recently it was found that these results are model depended,
i.e., depending on the specific geometry of space-time the results
may vary as the presence of non-diagonal components of
energy-momentum tensor of the spinor field imposes severe
restrictions to the space-time geometry as well
\cite{Saha2015CJP,Saha2015CnJP,Saha2016IJTP,Saha2016EPJP}.

Recently we have extended our study to Bianchi type- $VIII$ and $IX$
models \cite{SahaArXiV2017}. The main purpose of this paper is to
study the role of spinor field in the evolution of a Bianchi
type-$II$ space-time, as unlike Bianchi type- $VIII$ and $IX$ models
here we can exploit two different scenarios.

Bianchi type-$II$ model in presence of perfect fluid and anisotropic
dark energy was studied in \cite{Suresh}, whereas such a model with
massive string was considered in \cite{Singh}. Exact solutions to
$BII$ cosmological model was found in \cite{Reddy}, whereas a
similar model in presence of matter and electromagnetic field was
studied in \cite{Reddy1}. Exact solutions for Bianchi type-$II$
cosmological model in the Jordan Brans-Dicke scalar-tensor theory of
gravitation are obtained in \cite{Chauvet}. Some remarks on the
properties of Bianchi type-$II$ models were made in
\cite{SahaCEJP2011,SahaGnC2013}.  Magnetic Bianchi type-$II$ string
cosmological model in loop quantum cosmology was investigated in
\cite{SahaAPSS2014}. Nonlinear spinor field in Bianchi type-$II$
space-time was studied in \cite{SahaONP2014}

\section{Basic equation}

Let us consider the spinor field Lagrangian

\begin{equation}
L = \frac{\imath}{2} \biggl[\bp \gamma^{\mu} \nabla_{\mu} \psi-
\nabla_{\mu} \bar \psi \gamma^{\mu} \psi \biggr] - m_{\rm sp} \bp
\psi - F, \label{lspin}
\end{equation}
where the nonlinear term $F$ describes the self-action of a spinor
field and can be presented as some arbitrary functions of invariants
generated from the real bilinear forms of a spinor field. We
consider the case when $F = F(K)$ with $K$ taking one of the
followings values $\{I,\,J,\,I + J,\,I - J\}$.

Here $\nabla_{\mu}$ covariant derivative of the spinor field having
the form

\begin{subequations}
\begin{eqnarray}
\nabla_{\mu} \psi &=& \partial_\mu \psi - \G_\mu \psi,\\
\nabla_{\mu} \bp &=& \partial_\mu \bp + \bp \G_\mu,
\end{eqnarray}
\end{subequations}
where $\G_\mu$ is the spinor affine connection defined as

\begin{equation}
\Gamma_\mu = \frac{1}{4} \bg_{a} \gamma^\nu \partial_\mu e^{(a)}_\nu
- \frac{1}{4} \gamma_\rho \gamma^\nu \Gamma^{\rho}_{\mu\nu},
\label{sfc}
\end{equation}
where $\bg_a$ are the Dirac matrices in flat space-time,
$\gamma_\nu$ are the Dirac matrices in curved space-time,
$e^{(a)}_\nu$ are the tetrad and $\Gamma^{\rho}_{\mu\nu}$ are the
Christoffel symbols.

The corresponding energy momentum tensor is found from
\begin{eqnarray}
T_{\mu}^{\,\,\,\rho} &=& \frac{\imath}{4} g^{\rho\nu} \biggl(\bp
\gamma_\mu \nabla_\nu \psi + \bp \gamma_\nu \nabla_\mu \psi -
\nabla_\mu \bar \psi \gamma_\nu \psi - \nabla_\nu \bp \gamma_\mu
\psi \biggr) \,- \delta_{\mu}^{\rho} L \nonumber\\
&=&\frac{\imath}{4} g^{\rho\nu} \biggl(\bp \gamma_\mu
\partial_\nu \psi + \bp \gamma_\nu \partial_\mu \psi -
\partial_\mu \bar \psi \gamma_\nu \psi - \partial_\nu \bp \gamma_\mu
\psi \biggr)\nonumber\\
& - &\frac{\imath}{4} g^{\rho\nu} \bp \biggl(\gamma_\mu \G_\nu +
\G_\nu \gamma_\mu + \gamma_\nu \G_\mu + \G_\mu \gamma_\nu\biggr)\psi
 \,- \delta_{\mu}^{\rho} L. \label{temsp0}
\end{eqnarray}

Bianchi type $II$ space-time ($BII$) is given by

\begin{equation}
ds^2 =  dt^2 - a_1^2(t) dx_1^2 - [h^2(x_3) a_1^2(t) + f^2(x_3)
a_2^2(t)] dx_2^2 -  a_3^2 dx_3^2 + 2 a_1^2(t)h(x_3)dx_1 dx_2,
\label{bii-ix}
\end{equation}
with $a_1,\,a_2,\,a_3$ being the functions of $t$ and $h$ and $f$
are the functions of $x_3$ only, wherewith $f^{\prime \prime} = 0$.
As one sees, $f^{\prime \prime} = 0$ may lead to two different cases
with $f^{\prime} \ne 0$ and $f^{\prime} = 0$. We first consider the
case with $f^{\prime} \ne 0$ and later study the case with
$f^{\prime} = 0$. So let us begin with $f^{\prime} \ne 0$.

The spinor affine connections \eqref{sfc} corresponding to
\eqref{bii-ix} read

\begin{subequations}
\label{sfcII}
\begin{eqnarray}
\G_1 &=&  \frac{1}{2} \dot a_1 \bg^1\bg^0 - \frac{1}{4} \frac{a_1^2
h^\prime}{ a_2 a_3 f} \bg^2\bg^3, \label{G1ii-ix}\\
\G_2 &=&   \frac{1}{2} f \dot a_2 \bg^2\bg^0 - \frac{1}{2} h \dot
a_1 \bg^1\bg^0 - \frac{1}{4}\frac{a_1 h^\prime}{a_3}\bg^1\bg^3 +
\frac{1}{2} \frac{a_2 f^\prime}{ a_3}  \bg^2\bg^3 +
\frac{1}{4}\frac{a_1^2 h h^\prime}{a_2 a_3 f}\bg^2\bg^3,
\label{G2ii-ix}\\
\G_3 &= &  \frac{1}{2} \dot a_3 \bg^3\bg^0 + \frac{1}{4} \frac{a_1
h^\prime}{a_2 f} \bg^1 \bg^2, \label{G3ii-ix}\\
\G_0 &= &  0.\label{G0ii-ix}
\end{eqnarray}
\end{subequations}

In view of \eqref{sfcII}  the spinor field equations corresponding
to the Lagrangian \eqref{lspin} has the form

\begin{subequations}
\label{speq}
\begin{eqnarray}
\imath\gamma^\mu \nabla_\mu \psi - m_{\rm sp} \psi - {\cD}
\psi -  \imath {\cG} \gamma^5 \psi &=&0, \label{speq1} \\
\imath \nabla_\mu \bp \gamma^\mu +  m_{\rm sp} \bp + {\cD} \bp +
\imath {\cG}  \bp \gamma^5 &=& 0, \label{speq2}
\end{eqnarray}
\end{subequations}
where ${\cD} = 2 S F_K K_I$ and ${\cG} = 2 P F_K K_J$.

The system \eqref{speq} can be rewritten as
\begin{subequations}
\label{speqIIcom}
\begin{eqnarray}
\imath\bg^0 \dot \psi + \frac{\imath}{2}  \frac{\dot V}{V} \bg^0
\psi + \frac{1}{4} \frac{a_1 h^\prime}{ a_2 a_3 f}\bg^5 \bg^0 \psi +
\frac{\imath}{2} \frac{f^\prime}{ a_3 f} \bg^3 \psi - \left[m_{\rm
sp} +{\cD}\right]
\psi -  \imath {\cG} \bg^5 \psi &=&0, \label{speq1IIcom} \\
\imath \dot \bp \bg^0 + \frac{\imath}{2} \frac{\dot V}{V} \bp \bg^0
-\frac{1}{4}\frac{a_1 h^\prime}{ a_2 a_3 f} \bp \bg^5 \bg^0 +
\frac{\imath}{2} \frac{f^\prime}{ a_3 f} \bp \bg^3 + \left[m_{\rm
sp} + {\cD}\right] \bp +  \imath {\cG}  \bp \bg^5 &=& 0.
\label{speq2IIcom}
\end{eqnarray}
\end{subequations}

In view of \eqref{speq} the spinor field Lagrangian can be written
as

\begin{eqnarray}
L & = & \frac{\imath}{2} \bigl[\bp \gamma^{\mu} \nabla_{\mu} \psi-
\nabla_{\mu} \bar \psi \gamma^{\mu} \psi \bigr] - m_{\rm sp} \bp
\psi - F(K)
\nonumber \\
& = & \frac{\imath}{2} \bp [\gamma^{\mu} \nabla_{\mu} \psi - m_{\rm
sp} \psi] - \frac{\imath}{2}[\nabla_{\mu} \bar \psi \gamma^{\mu} +
m_{\rm sp} \bp] \psi
- F(K),\nonumber \\
& = & 2 (I F_I + J F_J) - F = 2 K F_K - F(K). \label{lspin01II}
\end{eqnarray}

The nontrivial components of Einstein tensor corresponding to
\eqref{bii-ix} are

\begin{subequations}
\label{EEBII_IX}
\begin{eqnarray}
G^1_1 &=& - \left(\frac{\ddot a_2}{a_2} + \frac{\ddot a_3}{a_3} +
\frac{\dot a_2}{a_2}\frac{\dot a_3}{a_3}\right) + \frac{a_1^2
h^2}{a_2^2 a_3^2 f^2}\left(\frac{3}{4}\frac{h^{\prime 2}}{h^2}
 + \frac{1}{2}\frac{h^{\prime \prime}}{h}
 - \frac{1}{2}\frac{h^\prime}{h}\frac{f^\prime}{f}\right),\label{EE11}\\
 G^2_1 &=& \frac{1}{2} \frac{a_1^2 h}{a_2^2 a_3^2
 f^2} \left(\frac{h^{\prime \prime}}{h} - \frac{h^\prime}{h}\frac{f^\prime}{f}\right),
 \label{EE12}\\
 G^1_2 &=& h \left(\frac{\ddot a_2}{a_2} - \frac{\ddot a_1}{a_1} +
\frac{\dot a_2}{a_2}\frac{\dot a_3}{a_3} - \frac{\dot
a_1}{a_1}\frac{\dot a_3}{a_3}\right) +
\frac{h}{a_3^2}\left(\frac{1}{2}\frac{h^{\prime\prime}}{h}  -
\frac{1}{2} \frac{h^\prime}{h}\frac{f^\prime}{f}\right) \nonumber\\
& &  + \frac{a_1^2 h^3 }{a_2^2 a_3^2 f^2} \left(\frac{1}{2}
\frac{h^\prime}{h}\frac{f^\prime}{f}  - \frac{1}{2}\frac{
h^{\prime\prime}}{h} - \frac{h^{\prime 2}}{h^2}\right),
\label{EE21}\\
G_2^2 &=& - \left(\frac{\ddot a_3}{a_3} + \frac{\ddot a_1}{a_1} +
\frac{\dot a_3}{a_3}\frac{\dot a_1}{a_1}\right) - \frac{1}{2}
\frac{a_1^2 h^2}{a_2^2 a_3^2 f^2} \left(\frac{ h^{\prime\prime}}{h}
- \frac{h^\prime}{h}\frac{f^\prime}{f} + \frac{1}{2}
\frac{h^{\prime 2}}{h^2}\right), \label{EE22}\\
G^3_3 &=& - \left(\frac{\ddot a_1}{a_1} + \frac{\ddot a_2}{a_2} +
\frac{\dot a_1}{a_1}\frac{\dot a_2}{a_2}\right) -
\frac{1}{4}\frac{a_1^2 h^2}{a_2^2 a_3^2 f^2}
\frac{ h^{\prime 2}}{h^2}, \label{EE33}\\
G^0_3 &=& \left(\frac{\dot a_2}{a_2} - \frac{\dot a_3}{a_3}\right) \frac{f^\prime}{f}, \label{EE30}\\
G^3_0 &=& - \frac{1}{a_3^2}\left(\frac{\dot a_2}{a_2} - \frac{\dot a_3}{a_3}\right) \frac{f^\prime}{f}, \label{EE03}\\
G^0_0 &=& -\left(\frac{\dot a_1}{a_1}\frac{\dot a_2}{a_2} +
\frac{\dot a_2}{a_2}\frac{\dot a_3}{a_3} + \frac{\dot
a_3}{a_3}\frac{\dot a_1}{a_1} \right) + \frac{1}{4}\frac{a_1^2
 h^2}{a_2^2 a_3^2 f^2}\frac{h^{\prime 2}}{h^2}.\label{EE00}
\end{eqnarray}
\end{subequations}
From \eqref{EEBII_IX} one finds
\begin{equation}
G^1_2 = h \left(G^2_2 - G^1_1\right) + \left(h^2 + \frac{a_2^2
f^2}{a_1^2}\right) G^2_1. \label{G2112}
\end{equation}

From \eqref{temsp0} one finds the following nontrivial components of
energy-momentum tensor

\begin{subequations}
\label{TEVBII_IX}
\begin{eqnarray}
T^{0}_{0} & = &   m_{\rm sp} S + F(K) -  \frac{1}{4} \frac{a_1
h^\prime}{a_2 a_3 f}A^0, \label{ApT00f}\\
T^{1}_{1} & = & \left[ F(K) - 2 K F_K\right] +
\frac{1}{4}\frac{h}{a_2^2 f} \left(a_1 \dot a_2 - \dot a_1
a_2\right) A^3 + \frac{1}{4}\frac{a_1 h}{a_2 a_3
f}\left(\frac{f^\prime}{f} - \frac{h^\prime}{h}\right)
A^0, \label{ApT11f}\\
T^{2}_{2} & = & \left[ F(K) - 2 K F_K\right] -
\frac{1}{4}\frac{h}{a_2^2 f} \left(a_1 \dot a_2 - \dot a_1
a_2\right) A^3 - \frac{1}{4}\frac{a_1 h}{a_2 a_3
f}\left(\frac{f^\prime}{f} - \frac{h^\prime}{h}\right) A^0,
\label{ApT22f}\\
T^{3}_{3} & = & \left[ F(K) - 2 K F_K\right] + \frac{1}{4}\frac{a_1
h}{a_2 a_3 f} \frac{h^\prime}{h} A^0, \label{ApT33f}\\
T^{1}_{2} & = & \frac{1}{4} \left(\frac{f}{a_1^2} - \frac{h^2}{a_2^2
f}\right) \left( a_1 \dot a_2 - \dot a_1 a_2\right) A^3 +
\left[\frac{a_1 h^2}{a_2 a_3 f} \left(\frac{1}{2}\frac{h^\prime}{h}
- \frac{1}{4}\frac{f^\prime}{f}\right) + \frac{1}{4}\frac{a_2 f}{a_1
a_3}\frac{f^\prime}{f}\right] A^0,\label{ApT12f}\\
T^{2}_{1} & = & \frac{1}{4}\frac{1}{a_2^2 f} \left( a_1 \dot a_2 -
\dot a_1 a_2\right) A^3 + \frac{1}{4}\frac{a_1}{a_2 a_3
f}\frac{f^\prime}{f} A^0,\label{ApT21f}\\
T^{2}_{3} & = & - \frac{1}{4}\frac{1}{a_2^2 f} \left(\dot a_2 a_3 -
a_2 \dot a_3\right) A^1,  \label{ApT23f}\\
T^{3}_{2} & = & - \frac{1}{4} \frac{1}{a_3^2} \left[ h\left(\dot a_1
a_3 - a_1 \dot a_3\right)A^2 + f \left(\dot a_2 a_3 - a_2 \dot
a_3\right)A^1 \right],\label{ApT32f}\\
T^{1}_{3} & = & \frac{1}{4}\frac{1}{a_1^2 }  \left(  \dot a_1 a_3 -
a_1 \dot a_3 \right) A^2 - \frac{1}{4} \frac{h}{a_2^2 f} \left(\dot
a_2 a_3 - a_2 \dot a_3\right) A^1,  \label{ApT13f}\\
T^{3}_{1} & = & \frac{1}{4} \frac{1}{a_3^2} \left( \dot a_1 a_3 -
a_1 \dot a_3 \right) A^2, \label{ApT31f}\\
T^{0}_{1} & = & \frac{1}{4} \left[ \frac{a_1^2 h^\prime}{ a_2 a_3
f} A^1 +  \frac{a_1 f^\prime}{a_3}  A^2\right], \label{ApT01f}\\
T^{1}_{0} & = & \frac{1}{4}\left[\left( \frac{a_1 h f^\prime}{a_2^2
a_3 f^2} + \frac{h f^\prime}{a_2 a_3 f} - \frac{h^\prime}{a_2 a_3
f}\right) A^1 -  \frac{f^\prime}{a_1 a_3 }  A^2\right],
\label{ApT10f}\\
T^{0}_{2} & = & - \frac{1}{4}\left[\left(\frac{a_1 f^\prime}{a_3} +
\frac{a_2 f f^\prime}{a_3} +  \frac{a_1^2 h h^\prime}{ a_2 a_3 f}
\right) A^1 + \frac{a_1 h f^\prime}{a_3}\,A^2\right],
\label{ApT02f}\\
T^{2}_{0} & = & \frac{1}{4}\left(\frac{f\prime}{a_2 a_3 f} +
\frac{a_1 f\prime}{a_2^2 a_3 f^2}\right) A^1,  \label{ApT20f}
\end{eqnarray}
\end{subequations}

From \eqref{TEVBII_IX} it can be shown that

\begin{equation}
T^1_2 = h \left(T^2_2 - T^1_1\right) + \left(h^2 + \frac{a_2^2
f^2}{a_1^2}\right) T^2_1. \label{T2112}
\end{equation}

Moreover from \eqref{TEVBII_IX} it follows that
\begin{equation}
T^3_2 = \frac{a_2^2 f^2}{a_3^2} T^2_3 - h T^3_1, \label{T322331}
\end{equation}

\begin{equation}
T^1_3 =  h T^2_3  + \frac{a_3^2}{a_1^2} T^3_1, \label{T132331}
\end{equation}

\begin{equation}
T^1_0 =  h T^2_0  - \frac{1}{a_1^2} T^0_1, \label{T102001}
\end{equation}

and

\begin{equation}
T^0_2 = - a_2^2 f^2 T^2_0  - h  T^0_1. \label{T022001}
\end{equation}

Then the system of Einstein equations

\begin{equation}
G^\mu_\nu = - \kappa T^\mu_\nu, \label{EE}
\end{equation}

on account of linearly dependent components takes the form
\begin{subequations}
\label{EECom}
\begin{eqnarray}
\left(\frac{\ddot a_2}{a_2} + \frac{\ddot a_3}{a_3} + \frac{\dot
a_2}{a_2}\frac{\dot a_3}{a_3}\right) &-&\frac{1}{2} \frac{a_1^2
h^2}{a_2^2 a_3^2 f^2}\left(\frac{h^{\prime \prime}}{h}
 - \frac{h^\prime}{h}\frac{f^\prime}{f} + \frac{3}{2}\frac{h^{\prime 2}}{h^2}\right)
  \label{EECom11}
 \\
 &=&\kappa \left[ \left( F(K) - 2 K F_K\right) +
\frac{1}{4}\frac{a_1 h}{a_2 f} \left(\frac{\dot a_2}{a_2} - \frac{
\dot a_1}{ a_1}\right) A^3 + \frac{1}{4}\frac{a_1 h}{a_2 a_3
f}\left(\frac{f^\prime}{f} - \frac{h^\prime}{h}\right) A^0\right],
  \nonumber\\
\left(\frac{\ddot a_3}{a_3} + \frac{\ddot a_1}{a_1} + \frac{\dot
a_3}{a_3}\frac{\dot a_1}{a_1}\right) & + &\frac{1}{2} \frac{a_1^2
h^2}{a_2^2 a_3^2 f^2} \left(\frac{ h^{\prime\prime}}{h} -
\frac{h^\prime}{h}\frac{f^\prime}{f} + \frac{1}{2} \frac{h^{\prime
2}}{h^2}\right) \label{EECom22}\\
&=& \kappa \left[ \left(F(K) - 2 K F_K\right) - \frac{1}{4}\frac{a_1
h}{a_2 f} \left(\frac{ \dot a_2}{a_2} - \frac{\dot a_1}{ a_1}\right)
A^3 - \frac{1}{4}\frac{a_1 h}{a_2 a_3 f}\left(\frac{f^\prime}{f} -
\frac{h^\prime}{h}\right) A^0\right],   \nonumber \\
 \left(\frac{\ddot a_1}{a_1} + \frac{\ddot a_2}{a_2} +
\frac{\dot a_1}{a_1}\frac{\dot a_2}{a_2}\right) & + &
\frac{1}{4}\frac{a_1^2 h^{\prime 2}}{a_2^2 a_3^2 f^2}  = \kappa
\left[ \left(F(K) - 2 K F_K\right) + \frac{1}{4}\frac{a_1
h^\prime}{a_2 a_3 f}  A^0\right], \label{EECom33}\\
\left(\frac{\dot a_1}{a_1}\frac{\dot a_2}{a_2} + \frac{\dot
a_2}{a_2}\frac{\dot a_3}{a_3} + \frac{\dot a_3}{a_3}\frac{\dot
a_1}{a_1} \right) & - & \frac{1}{4}\frac{a_1^2
 h^{\prime 2}}{a_2^2 a_3^2 f^2}  =\kappa \left[ m_{\rm sp}
S + F(K) -  \frac{1}{4} \frac{a_1
h^\prime}{a_2 a_3 f}A^0\right], \label{EECom00}\\
\frac{1}{2} \frac{a_1^2 h}{a_2^2 a_3^2
 f^2} \left(\frac{h^{\prime \prime}}{h} -
 \frac{h^\prime}{h}\frac{f^\prime}{f}\right) &=& -\kappa \frac{1}{4}
 \frac{1}{a_2 f}   \left[\left( \frac{ \dot a_2}{a_2} -
\frac{\dot a_1}{ a_1}\right) A^3 + \frac{a_1}{ a_3
}\frac{f^\prime}{f} A^0\right], \label{EECom21}\\
\left(\frac{\dot a_2}{a_2} - \frac{\dot a_3}{a_3}\right)
\frac{f^\prime}{f} &=& 0, \label{EECom03}\\
0 & =&\left(\frac{\dot a_2}{a_2} -
\frac{\dot a_3}{a_3}\right) A^1, \label{EECom23}\\
0 & =&  \left( \frac{\dot a_1}{a_1} -
\frac{\dot a_3}{a_3} \right) A^2, \label{EECom31}\\
0 & =&  \left[ \frac{a_1 h^\prime}{ a_2
f} A^1 +   f^\prime  A^2\right], \label{EECom01}\\
0 & =& \left(1 + \frac{a_1}{a_2 f}\right) \frac{f^\prime}{f}\, A^1.
\label{EECom20}
\end{eqnarray}
\end{subequations}

The summation of \eqref{EECom11}, \eqref{EECom22}, \eqref{EECom33}
and 3 times \eqref{EECom00} gives

\begin{eqnarray}
\frac{\ddot V}{V} - \frac{1}{2} \frac{a_1^2
 h^{\prime 2}}{a_2^2 a_3^2 f^2}  = \frac{3
 \kappa}{2}\left[m_{\rm sp} S + 2 \left(F - K F_K\right) -
 \frac{1}{2}  \frac{a_1 h^\prime}{a_2 a_3 f}A^0\right], \label{EqV}
\end{eqnarray}

where
\begin{equation}
V = a_1 a_2 a_3 \label{Vdef-II-IX}
\end{equation}
is the volume scale.

 Then in view of $\frac{f^\prime}{f} \ne 0$ from
\eqref{EECom03} we find $\left(\frac{\dot a_2}{a_2} - \frac{\dot
a_3}{a_3}\right) = 0$, on the other hand for same reason
\eqref{EECom20} yields $A^1 = 0$, whereas inserting $A^1 = 0 $ into
\eqref{EECom01} we obtain $A^2 = 0$. Thus if we consider
$\frac{f^\prime}{f} \ne 0$ from \eqref{EECom03} - \eqref{EECom20} we
have

\begin{eqnarray}
A^1 = 0, \quad A^2 = 0, \quad \left(\frac{\dot a_2}{a_2} -
\frac{\dot a_3}{a_3}\right) = 0. \label{03-20}
\end{eqnarray}
In view of $A^2 = 0$ the equation \eqref{EECom31} yields two
possibilities:

\begin{eqnarray}
\left(\frac{\dot a_1}{a_1} - \frac{\dot a_3}{a_3}\right) \ne 0,
\label{3-1ne0}
\end{eqnarray}
which means $a_2 \sim a_3$, or
\begin{eqnarray}
\left(\frac{\dot a_1}{a_1} - \frac{\dot a_3}{a_3}\right) = 0,
\label{3-1eq0}
\end{eqnarray}
which means $a_1 \sim a_2 \sim a_3$.

It can be shown that the spinor field invariants in this case obey
the following equations:

\begin{subequations}
\label{SpinInv}
\begin{eqnarray}
\dot S_0   + \frac{1}{2} \frac{a_1 h^\prime}{ a_2 a_3 f} P_0  + 2
{\cG} A^0_0 & = & 0, \label{S0Inv}\\
\dot P_0   + \frac{1}{2} \frac{a_1 h^\prime}{ a_2 a_3 f} S_0  - 2
\left[m_{\rm sp} +{\cD}\right] A^0_0 &=& 0, \label{P0Inv}\\
 \dot A^0_0   +
\frac{1}{2} \frac{f^\prime}{a_3 f} A^3_0  + 2 \left[m_{\rm sp}
+{\cD}\right] P_0 + 2 {\cG} S_0 &=& 0, \label{A00Inv}\\
 \dot A^3_0   +
\frac{1}{2} \frac{f^\prime}{a_3 f} A^0_0  &=& 0, \label{A30Inv}\\
\dot A^2_0 &=& 0. \label{A20Inv}
\end{eqnarray}
\end{subequations}
From \eqref{SpinInv} it can be easily shown that

\begin{equation}
P_0^2 - S_0^2 + \left(A_0^0\right)^2 - \left(A_0^3\right)^2  =
\tilde{c_1}, \quad  A^2_0 = \tilde{c_2}, \quad \tilde{c_1},\,
\tilde{c_2} - {\rm const.}, \label{InVCons}
\end{equation}
On account of \eqref{03-20} in this case we have $\tilde{c_2} = 0$.
On the other hand from Fierz theorem we have

\begin{equation}
I_A = \left(A^0\right)^2 - \left(A^1\right)^2 - \left(A^2\right)^2 -
\left(A^3\right)^2 = -\left(S^2 + P^2\right), \label{InVCons1}
\end{equation}

Now taking into account that $A^1 = 0$ and $A^2 = 0$ from
\eqref{InVCons1} one finds

\begin{equation}
\left(A_0^0\right)^2 - \left(A_0^3\right)^2 = -\left(S_0^2 +
P_0^2\right), \label{InVCons2}
\end{equation}

Then inserting \eqref{InVCons2} into \eqref{InVCons} one finds
\begin{equation}
S^2 = \frac{\rm const}{V^2}, \label{InVS0}
\end{equation}

Let us now impose the proportional condition between the shear and
expansion. Assuming that

\begin{equation}
\sigma_1^1 = q_1 \vartheta, \quad q_1 = {\rm const.} \label{propcon}
\end{equation}
for the metric function we have \cite{SahaArXiV2017}
\begin{equation}
a_i = X_i V^{Y_i}, \quad \prod_{i = 1}^3 X_i  = 1, \quad \sum_{i =
1}^3 Y_i = 0.  \label{a123Vexp}
\end{equation}
In this concrete case we have $X_1 = q_2$,\,\,\,$X_2 =
\sqrt{{q_3}/{q_2}}$,\,\,\, $X_3 = {1}/{\sqrt{q_2 q_3}}$,\,\,\,$ Y_1
= q_1 + 1/3,$\,\, and \,\,$Y_2 = Y_3 = 1/3 - q_1/2.$ Here $q_2$ and
$q_3$ are arbitrary constants. On account of \eqref{a123Vexp} the
equation for $V$ \eqref{EqV}  can be rewritten as

\begin{eqnarray}
\frac{\ddot V}{V} - \frac{1}{2} \frac{
 h^{\prime 2}}{f^2} q_2^4 V^{4q_1 - 2/3}   = \frac{3
 \kappa}{2}\left[m_{\rm sp} S + 2 \left(F - K F_K\right) -
 \frac{q_2^2}{2}  \frac{h^\prime}{f}  V^{2q_1 - 1/3} A^0\right], \label{EqV1}
\end{eqnarray}

Let us now rewrite \eqref{EECom21}. In view of \eqref{A30Inv}  this
equations can be written as

\begin{eqnarray}
 \frac{1}{f} \left(h^{\prime \prime} -
 \frac{f^\prime}{f} h^\prime\right) = -\frac{\kappa}{2}
 \frac{a_2 a_3^2}{a_1^2}   \left[\left( \frac{ \dot a_2}{a_2} -
\frac{\dot a_1}{ a_1}\right) A_0^3 - 2 a_1 {\dot A_0^3}\right],
\label{EECom21n1}
\end{eqnarray}
Now the left hand side of \eqref{EECom21n1} depends of $x_3$ only,
while the right hand side depends on only $t$. So we can finally
write the following system
\begin{subequations}
\label{eqha3}
\begin{eqnarray}
\frac{1}{f} \left(h^{\prime \prime} -
 \frac{f^\prime}{f} h^\prime\right) &=& b, \label{eqh0}\\
\frac{\kappa}{2}
 \frac{a_2a_3^2}{a_1^2}   \left[\left( \frac{ \dot a_2}{a_2} -
\frac{\dot a_1}{ a_1}\right) A_0^3 - 2 a_1 {\dot A_0^3}\right] &=& -
b, \label{eqa30}
\end{eqnarray}
\end{subequations}
Inserting \eqref{a123Vexp} one finally finds
\begin{subequations}
\label{eqha3n}
\begin{eqnarray}
h^{\prime \prime} -
 \frac{f^\prime}{f} h^\prime &=& b f, \label{eqh}\\
\dot A_0^3 + \frac{3 q_1}{4 q_2} \frac{\dot V}{V^{q_1 + 4/3}} A_0^3
&=& \frac{b}{\kappa} \sqrt{q_2^5 q_3}V^{5q_1/2 - 2/3}, \label{eqa3}
\end{eqnarray}
\end{subequations}

Exploiting \eqref{A30Inv} we rewrite \eqref{EqV1} in the following
way

\begin{eqnarray}
\ddot V &-& \frac{3 \kappa}{2} \sqrt{\frac{q_2^3}{q_3}}
\frac{h^\prime}{f^\prime}\dot A_0^3 V^{3q_1/2} - \frac{1}{4} q_2^4
V^{4q_1 + 1/3}\left(\frac{h^\prime}{f}\right)^2  = \frac{3
 \kappa}{2}\left[m_{\rm sp} S + 2 \left(F - K F_K\right)\right] V, \label{EqVBIInew}
\end{eqnarray}

To find the solution to the equation \eqref{EqVBIInew} we have to
give the concrete form of spinor field nonlinearity. Following some
previous papers, we choose the nonlinearity to be the function of
$S$ only, having the form
\begin{equation}
F = \sum_{k} \lambda_k I^{n_k} =  \sum_{k} \lambda_k S^{2 n_k}.
\label{nonlinearity}
\end{equation}
Recently, this type of nonlinearity was considered in a number of
papers \cite{Saha2015CJP,Saha2015CnJP,Saha2016IJTP,Saha2016EPJP}.
For simplicity we consider only three terms of the sum. We set $n_k
= n_0: 1 - 2 n_0 = 0$ which gives $n_0 = 1/2$. In this case the
corresponding term can be added with the mass term. We assume that
$q_1$ is a positive quantity, so that $4q_1 + 1/3$ is positive too.
For the nonlinear term to be dominant at large time, we set $n_k =
n_1: 1 -2 n_1 > 4q_1 + 1/3$, i.e., $n_1 < 1/3 - 2 q_1 $. And
finally, for the nonlinear term to be dominant at the early stage we
set $n_k = n_2: 1 - 2 n_2 < 0$, i.e., $n_2 > 1/2$. Then the spinor
field nonlinearity can be written as
\begin{equation}
F = \lambda_0 I^{n_0} +  \lambda_1 I^{n_1} +  \lambda_2 I^{n_2}.
\label{nonlinearity1}
\end{equation}

Once we have the concrete form of nonlinearity, we can solve the
foregoing equation. In doing so we rewrite the equations
\eqref{EqVBIInew} together with \eqref{eqa3} as follows

\begin{subequations}
\label{YVA3}
\begin{eqnarray}
\dot V & = & Y, \label{Vfin}\\
\dot Y & = & \frac{3 \kappa}{2} \sqrt{\frac{q_2^3}{q_3}}
\frac{h^\prime}{f^\prime}V^{3q_1/2} \Phi_1(V, A_0^3, Y) +
\Phi_2(V, A_0^3, Y), \label{Yfin}\\
\dot A_0^3 &=& \Phi_1(V, A_0^3, Y). \label{A03fin}
\end{eqnarray}
\end{subequations}

where we denote

\begin{eqnarray}
\Phi_1 (V, A_0^3, Y) &=& -\frac{3 q_1}{ 4 q_2}\,\,A_0^3\,\, V^{-(q_1
+ 4/3)} Y
+ \frac{b}{\kappa} \sqrt{q_2^5 q_3} V^{5q_1/2 - 2/3},\nonumber\\
\Phi_2(V, A_0^3, Y) &=&  \frac{1}{4} q_2^4 V^{4q_1 + 1/3}
\left(\frac{h^\prime}{f}\right)^2  \nonumber\\
 &+&  \frac{3\kappa}{2}
\left[\left(m_{\rm sp} + \lambda_0\right) + 2 \lambda_1( 1 - n_1)
V^{1 - 2n_1} + 2 \lambda_2( 1 - n_2) V^{1 - 2n_2} \right].\nonumber
\end{eqnarray}

Let us recall that by definition we have $f^{\prime\prime} = 0$.
Then as per our assumption $f^\prime \ne 0$ we find

\begin{equation}
f = c_1 x_3 + c_2, \quad c_1,\,\,c_2 - {\rm const.} \label{f1}
\end{equation}
Inserting this into \eqref{eqh} in case of $b \ne 0$ we find
\begin{equation}
h = \frac{1}{3} b c_1 x_3^3 + \frac{1}{2} \left(c_3 c_1 + b
c_2\right) x_3^2 + c_3 c_2 x_3 + c_4, \quad c_3,\,\,c_4 - {\rm
const.}, \label{h1}
\end{equation}
and in case of $b = 0$ we obtain
\begin{equation}
h = \frac{1}{2} c_5 c_1 x_3^2 + c_5 c_2 x_3, \quad c_5 - {\rm
const.} \label{h2}
\end{equation}

Let us now numerically solve the system \eqref{YVA3}. Since we are
interested in qualitative picture of evolution, let us set $q_2 =
1,\, q_3 = 1$ and $\kappa = 1$. We also assume $V_0 = 1$.

We set $m_{\rm sp} = 1$ and $l_0 = 2$. As far as $q_1$, $n_1$ and
$n_2$ are concerned, in line of our previous discussions we choose
them in such a way that the power of nonlinear term in the equations
become integer. We have also studied the case for some different
values, but they didn't give any principally different picture. We
choose $q_1 = 2/3$,\, $n_1 = -3/2 < 1/3 - 2 q-1 = -1$ and $n_2 = 3/2
> 1/2$. It should be noted that we have taken some others value for
$q_1$ such as $q_1 = -1$, but it does not give qualitatively
different picture. We have also set $x_3 = [0,\,1] = 0.2 k $ with
step $k = 0..5$. Finally we have considered time span $[0,\,2]$ with
step size $0.001$. Here we consider different values of $\lambda_i$
both positive and negative. We choose the initial values for $V(0) =
1$, $ Y(0) = \dot V(0) = 0.1$, and $A_0^3 (0) = 1$, respectively.

In  Fig. \ref{Fig01} we have plotted the phase diagram of $[V,\,
\dot V,\, A_0^3]$ for a trivial $b$ and positive $\lambda_1$ and
$\lambda_2$.  The corresponding evolution of $V$ is given in Fig.
\ref{Fig02}. As one sees, in this case we have oscillatory mode of
expansion.

In Fig. \ref{Fig03} the phase diagram of $[V,\, \dot V,\, A_0^3]$
has been plotted for a trivial $b$, positive $\lambda_1$ and
negative $\lambda_2$, while the evolution of $V$ corresponding to it
is illustrated in Fig. \ref{Fig04}.

In Figs. \ref{Fig05} and \ref{Fig06}  the phase diagram of $[V,\,
\dot V,\, A_0^3]$ and volume scale $V$ are manifested for a trivial
$b$, trivial $\lambda_1$ and positive $\lambda_2$.

In Figs. \ref{Fig07}, \ref{Fig09} and \ref{Fig11} we have
illustrated the phase diagram of $[V,\, \dot V,\, A_0^3]$ for
$\lambda_1$ and $\lambda_2$ considered in Figs. \ref{Fig01},
\ref{Fig03} and \ref{Fig05}, but with a nontrivial $b$. Evolution of
$V$ corresponding to Figs. \ref{Fig07}, \ref{Fig09} and \ref{Fig11}
are illustrated in Figs. \ref{Fig08}, \ref{Fig10} and \ref{Fig12}

In the Figures each color corresponds to a concrete value of $x_3 =
0.2 k$, namely, red, green, yellow, blue, magenta and black color
corresponds to $k = 0,1,2,3,4,5$

\myfigures{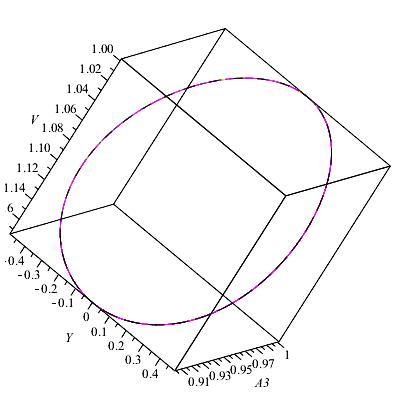}{0.40}{Phase diagram of $[V,\, \dot V,\, A_0^3]$ in
case of $b = 0$, $\lambda_1 = 1$ and $\lambda_2 =
1$}{0.45}{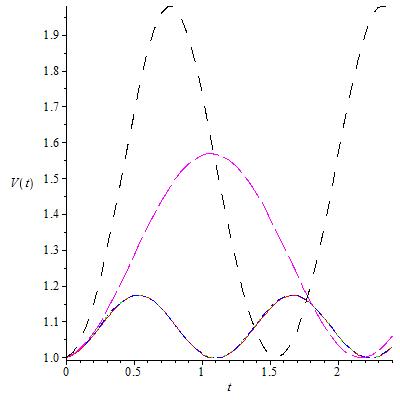}{0.40}{Evolution of $V$ corresponding to the phase
diagram given in Fig. \ref{Fig01} }{0.45}

\myfigures{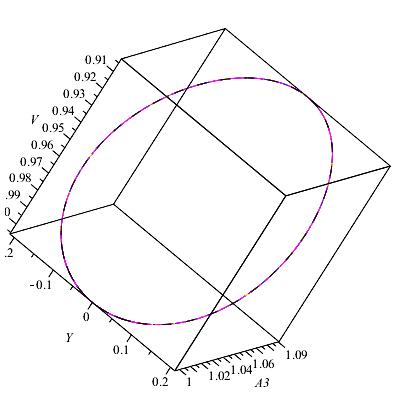}{0.40}{Phase diagram of $[V,\, \dot V,\, A_0^3]$ in
case of $b = 0$, $\lambda_1 = 1$ and $\lambda_2 = -0.0
1$}{0.45}{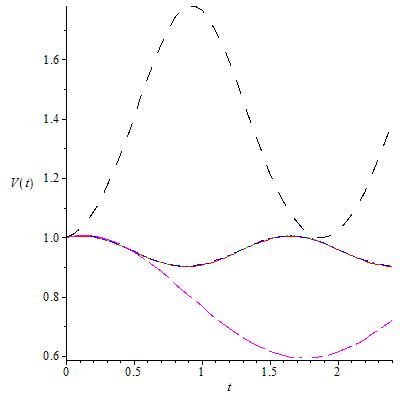}{0.40}{Evolution of $V$ corresponding to the phase
diagram given in Fig. \ref{Fig04} }{0.45}

\myfigures{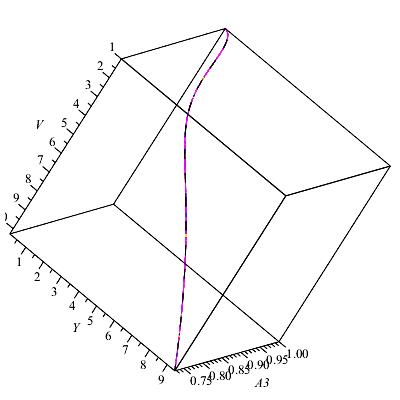}{0.40}{Phase diagram of $[V,\, \dot V,\, A_0^3]$ in
case of $b = 0$, $\lambda_1 = 0$ and $\lambda_2 =
1$}{0.45}{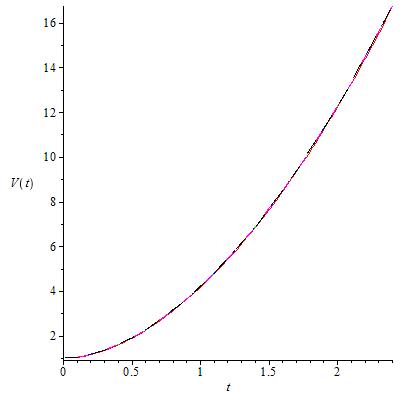}{0.40}{Evolution of $V$ corresponding to the phase
diagram given in Fig. \ref{Fig05} }{0.45}

\myfigures{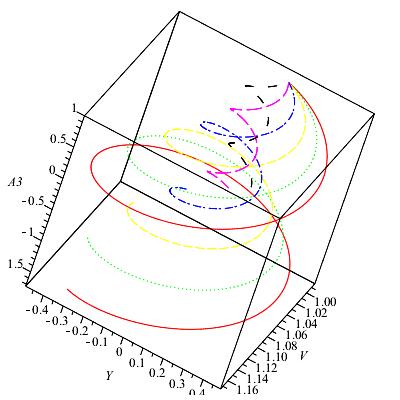}{0.40}{Phase diagram of $[V,\, \dot V,\, A_0^3]$ in
case of $b = 1$, $\lambda_1 = 1$ and $\lambda_2 =
1$}{0.45}{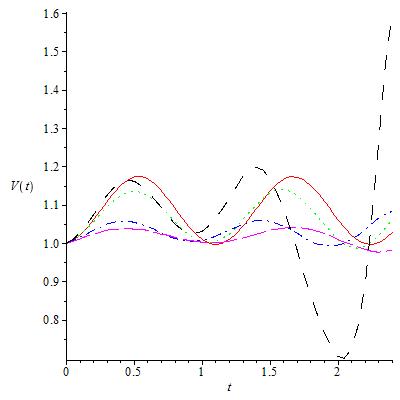}{0.40}{Evolution of $V$ corresponding to the phase
diagram given in Fig. \ref{Fig07} }{0.45}

\myfigures{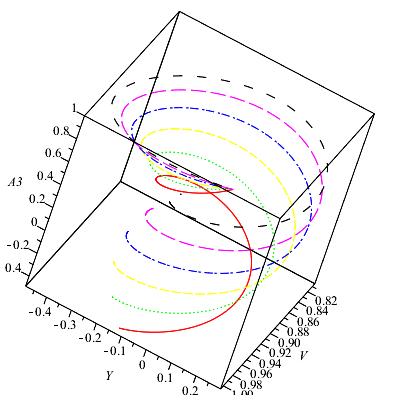}{0.40}{Phase diagram of $[V,\, \dot V,\, A_0^3]$ in
case of $b = 1$, $\lambda_1 = 1$ and $\lambda_2 = -0.0
1$}{0.45}{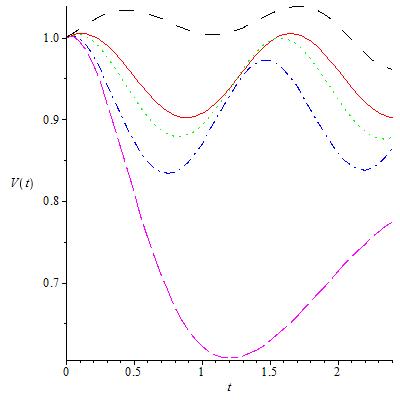}{0.40}{Evolution of $V$ corresponding to the phase
diagram given in Fig. \ref{Fig09} }{0.45}

\myfigures{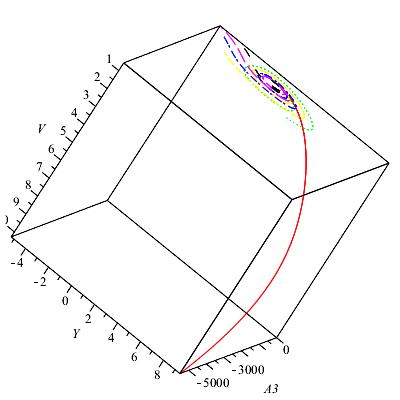}{0.40}{Phase diagram of $[V,\, \dot V,\, A_0^3]$ in
case of $b = 1$, $\lambda_1 = 0$ and $\lambda_2 =
1$}{0.45}{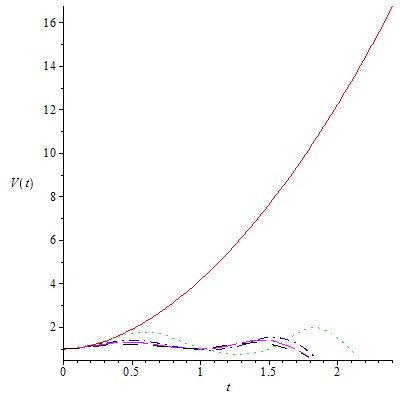}{0.40}{Evolution of $V$ corresponding to the phase
diagram given in Fig. \ref{Fig11} }{0.45}

\smallskip

Let us now consider the case when $f^\prime = 0$, i.e. $f = f_0 =
{\rm const.}$ In this case a few components of energy-momentum
tensor die out and finally we have the following system of Einstein
equations:

\begin{subequations}
\label{EEComf}
\begin{eqnarray}
\left(\frac{\ddot a_2}{a_2} + \frac{\ddot a_3}{a_3} + \frac{\dot
a_2}{a_2}\frac{\dot a_3}{a_3}\right) &-&\frac{1}{2} \frac{a_1^2
h^2}{a_2^2 a_3^2 f^2}\left(\frac{h^{\prime \prime}}{h}
  + \frac{3}{2}\frac{h^{\prime 2}}{h^2}\right)
  \label{EECom11f}
 \\
 &=&\kappa \left[ \left( F(K) - 2 K F_K\right) +
\frac{1}{4}\frac{a_1 h}{a_2 f} \left(\frac{\dot a_2}{a_2} - \frac{
\dot a_1}{ a_1}\right) A^3 - \frac{1}{4}\frac{a_1 h}{a_2 a_3 f}
\frac{h^\prime}{h} A^0\right],
  \nonumber\\
\left(\frac{\ddot a_3}{a_3} + \frac{\ddot a_1}{a_1} + \frac{\dot
a_3}{a_3}\frac{\dot a_1}{a_1}\right) & + &\frac{1}{2} \frac{a_1^2
h^2}{a_2^2 a_3^2 f^2} \left(\frac{ h^{\prime\prime}}{h}  +
\frac{1}{2} \frac{h^{\prime
2}}{h^2}\right) \label{EECom22f}\\
&=& \kappa \left[ \left(F(K) - 2 K F_K\right) - \frac{1}{4}\frac{a_1
h}{a_2 f} \left(\frac{ \dot a_2}{a_2} - \frac{\dot a_1}{ a_1}\right)
A^3 + \frac{1}{4}\frac{a_1 h}{a_2 a_3 f}
\frac{h^\prime}{h} A^0\right],   \nonumber \\
 \left(\frac{\ddot a_1}{a_1} + \frac{\ddot a_2}{a_2} +
\frac{\dot a_1}{a_1}\frac{\dot a_2}{a_2}\right) & + &
\frac{1}{4}\frac{a_1^2 h^{\prime 2}}{a_2^2 a_3^2 f^2}  = \kappa
\left[ \left(F(K) - 2 K F_K\right) + \frac{1}{4}\frac{a_1
h^\prime}{a_2 a_3 f}  A^0\right], \label{EECom33f}\\
\left(\frac{\dot a_1}{a_1}\frac{\dot a_2}{a_2} + \frac{\dot
a_2}{a_2}\frac{\dot a_3}{a_3} + \frac{\dot a_3}{a_3}\frac{\dot
a_1}{a_1} \right) & - & \frac{1}{4}\frac{a_1^2
 h^{\prime 2}}{a_2^2 a_3^2 f^2}  =\kappa \left[ m_{\rm sp}
S + F(K) -  \frac{1}{4} \frac{a_1
h^\prime}{a_2 a_3 f}A^0\right], \label{EECom00f}\\
\frac{1}{2} \frac{a_1^2 h}{a_2^2 a_3^2
 f^2} \frac{h^{\prime \prime}}{h}  &=& -\kappa \frac{1}{4}
 \frac{1}{a_2 f}  \left( \frac{ \dot a_2}{a_2} -
\frac{\dot a_1}{ a_1}\right) A^3, \label{EECom21f}\\
0 & =&\left(\frac{\dot a_2}{a_2} -
\frac{\dot a_3}{a_3}\right) A^1, \label{EECom23f}\\
0 & =&  \left( \frac{\dot a_1}{a_1} -
\frac{\dot a_3}{a_3} \right) A^2, \label{EECom31f}\\
0 & =&  \frac{a_1 h^\prime}{ a_2 f} A^1, \label{EECom01f}
\end{eqnarray}
\end{subequations}

The system for the invariants of bilinear spinor forms in this case
reads

\begin{subequations}
\label{SpinInvf}
\begin{eqnarray}
\dot S_0   + \frac{1}{2} \frac{a_1 h^\prime}{ a_2 a_3 f} P_0  + 2
{\cG} A^0_0 & = & 0, \label{S0Invf}\\
\dot P_0   + \frac{1}{2} \frac{a_1 h^\prime}{ a_2 a_3 f} S_0  - 2
\left[m_{\rm sp} +{\cD}\right] A^0_0 &=& 0, \label{P0Invf}\\
 \dot A^0_0   +  2 \left[m_{\rm sp}
+{\cD}\right] P_0 + 2 {\cG} S_0 &=& 0, \label{A00Invf}\\
 \dot A^3_0   &=& 0, \label{A30Invf}\\
 \dot A^2_0   &=& 0. \label{A20Invf}
\end{eqnarray}
\end{subequations}

From \eqref{S0Invf}, \eqref{P0Invf} and \eqref{A00Invf} we find
\begin{equation}
\left(P_0\right)^2 - \left(S_0\right)^2 + \left(A_0^0\right)^2 =
C_1^2, \quad C_1^2 = {\rm const.}, \label{Invf}
\end{equation}
whereas \eqref{A20Invf} and \eqref{A30Invf} yield
\begin{equation}
A_0^2 = C_2  \Rightarrow A^2 = \frac{C_2}{V}, \quad A_0^3 = C_3
\Rightarrow A^3 = \frac{C_3}{V}, \quad C_2, C_3 = {\rm const.}
\label{InvfA3}
\end{equation}
On the other hand from \eqref{InVCons1} on account of $A^1 = 0$
which follows from \eqref{EECom01f} and \eqref{InvfA3} we find

\begin{equation}
\left(P_0\right)^2 + \left(S_0\right)^2 + \left(A_0^0\right)^2 =
C_2^2 + C_3^2. \label{PSA3Invf}
\end{equation}

Then in this case we finally find

\begin{equation}
\left(S_0\right)^2 = \frac{1}{2}\left(C_2^2 + C_3^2 - C_1^2\right) =
C_{-}^2, \quad \left(P_0\right)^2 + \left(A_0^0\right)^2 =
\frac{1}{2}\left(C_2^2 + C_3^2 +  C_1^2\right) = C_{+}^2.
\label{SPA0}
\end{equation}
Then inserting $S_0$ and $P_0$ from \eqref{SPA0} into
\eqref{A00Invf} for $A_0^3$ we find

\begin{equation}
 \dot A^0_0   +  2 \left[m_{\rm sp}
+{\cD}\right] \sqrt{C_{+}^2- \left(A_0^0\right)^2}
 + 2 {\cG} C_{-} = 0. \label{A00Invfnew}\\
\end{equation}
The equation for $V$ in this case coincides with \eqref{EqV}, so to
solve it we again assume the proportionality condition that gives
$a_i = X_i V^{Y_i}, \quad \prod_{i = 1}^3 X_i  = 1, \quad \sum_{i =
1}^3 Y_i = 0$.

In what follows we solve the equation for $V$ numerically. In doing
so we consider the spinor field nonlinearity in the form
\eqref{nonlinearity1}. Taking into account that in this case ${\cG}
= 2 P F_K K_J = 0$, we write the equations  \eqref{EqV} and
\eqref{A00Invfnew} in the following way:

\begin{subequations}
\label{YVA0}
\begin{eqnarray}
\dot V & = & Y, \label{Vfin0}\\
\dot Y & = & \Phi_2(V, A_0^0) - \frac{3 q_2^2}{4}\frac{h^\prime}{f} V^{2 q_1 - 2/3} A_0^0, \label{Yfin0}\\
\dot A_0^0 &=& - 2 \left[\left(m_{\rm sp} + \lambda_0\right) + 2
\lambda_1 n_1 V^{1 - 2 n_1} + \lambda_2 n_2 V^{1 - 2 n_2}\right]
\sqrt{C_{+}^2 - \left(A_0^0\right)^2} . \label{A03fin0}
\end{eqnarray}
\end{subequations}

where we denote

\begin{eqnarray}
\Phi_2(V, A_0^0, Y) &=&  \frac{h^{\prime 2}}{2 f^2} q_2^4 V^{4q_1 +
1/3}
 \nonumber\\
 &+&  \frac{3\kappa}{2}
\left[\left(m_{\rm sp} + \lambda_0\right) + 2 \lambda_1( 1 - n_1)
V^{1 - 2n_1} + 2 \lambda_2( 1 - n_2) V^{1 - 2n_2} \right].\nonumber
\end{eqnarray}

To solve the foregoing system we have to find $h$ explicitly that
corresponds to $f = f_0 = {\rm const.}$ In doing so we rewrite the
equation \eqref{EECom21f} as

\begin{subequations}
\begin{eqnarray}
h^{\prime \prime}  & = & b f, \label{EECom21f0h}\\
\frac{\kappa C_3}{2}
 \frac{2 X_2 X_3^2}{X_1^2}  \left(Y_2 -
Y_1\right) V^{(Y_2 + 2 Y_3 - 2 Y_1 -2)}\,\, \dot V &=& b.
\label{EECom21f0A3}
\end{eqnarray}
\end{subequations}

In case of $b \ne 0$ on account of $f = f_0$ from \eqref{EECom21f0h}
one finds $h = b f_0 x_3^2/2 + c_1 x_3 + c_2.$ In this case equation
\eqref{EECom21f0A3} yields

\begin{equation}
V = \left[ \frac{ b X_1^2}{\kappa C_3 X_2 X_3^2 \left(Y_2 -
Y_1\right) \left(Y_2 + 2 Y_3 - 2 Y_1 - 1\right)}\, t +
t_0\right]^{1/(Y_2 + 2Y_3 - 2Y_1 - 1)}. \label{VA3}
\end{equation}
As one sees, in this case $V$ does not depend on spinor field,
whereas the equation \eqref{EqV} explicitly contains terms
corresponding to spinor field. This imposes some severe restrictions
on the choice of spinor field nonlinearity.

In case of $b = 0$ from \eqref{EECom21f0h} we find $h = c_1 x_3 +
c_2$ with $c_1 $ and $c_2$ being some arbitrary constants. Equation
\eqref{EECom21f0A3} in this case leads to $Y_1 = Y_2$, i.e. $a_1
\propto a_2$. As far as $V$ and $A_0^0$ ar concerned, in this case
we find it solving \eqref{YVA0} numerically. In doing so we set the
parameters as we did in previous case. For this we have to
substitute $h^\prime/f$ with $c_1/f_0$. As a result we see that
unlike the previous cases or $BVIII$ and $BIX$ models, the system
for defining $V$ does not depend on $x_3$ explicitly. In Figs.
\ref{Fig13}, \ref{Fig15}, \ref{Fig17} and \ref{Fig19} we have
plotted the pase diagram of $[V,\, \dot V,\, A_0^0]$ for $(\lambda_1
= -1$ and $\lambda_2 = -0.1)$, $(\lambda_1 = 1$ and $\lambda_2 =
1)$, $(\lambda_1 = 1$ and $\lambda_2 = -0.1)$ and $(\lambda_1 = -1$
and $\lambda_2 = 1)$, respectively with $b$ taken to be trivial.
Corresponding pictures of $V$ are given in Figs. \ref{Fig14},
\ref{Fig16}, \ref{Fig18} and \ref{Fig20}, respectively.

\myfigures{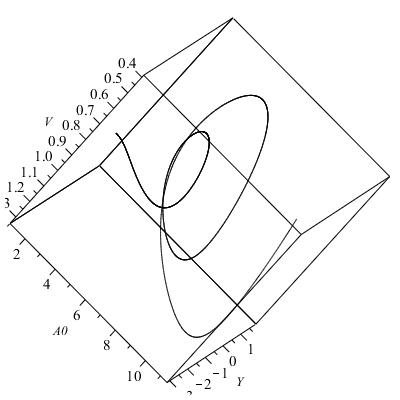}{0.40}{Phase diagram of $[V,\, \dot V,\, A_0^0]$ in
case of $b = 0$, $\lambda_1 = -1$ and $\lambda_2 =
-0.1$}{0.45}{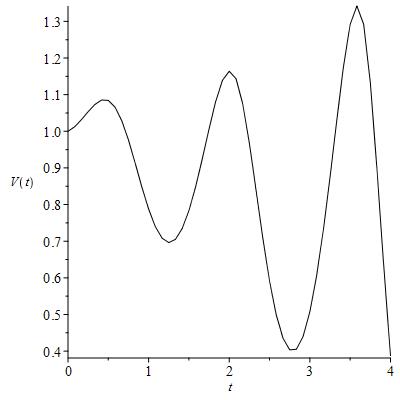}{0.40}{Evolution of $V$ corresponding to the
phase diagram given in Fig. \ref{Fig13} }{0.45}

\myfigures{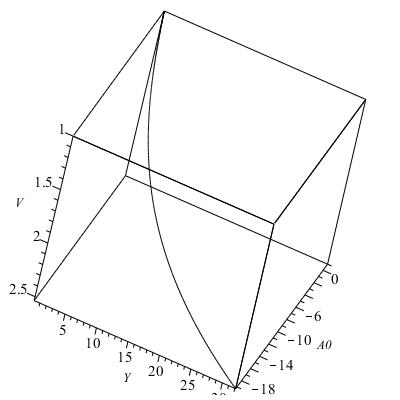}{0.40}{Phase diagram of $[V,\, \dot V,\, A_0^0]$ in
case of $b = 0$, $\lambda_1 = 1$ and $\lambda_2 =
1$}{0.45}{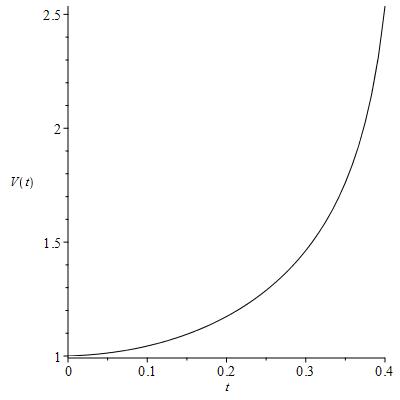}{0.40}{Evolution of $V$ corresponding to the phase
diagram given in Fig. \ref{Fig15} }{0.45}

\myfigures{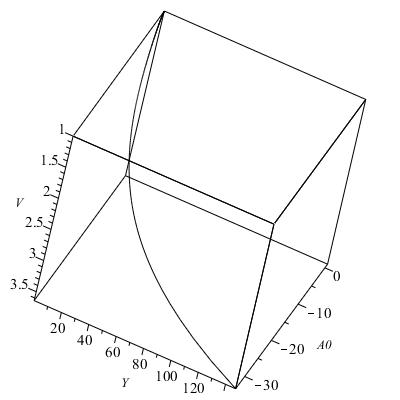}{0.40}{Phase diagram of $[V,\, \dot V,\, A_0^0]$ in
case of $b = 0$, $\lambda_1 = 1$ and $\lambda_2 =
-0.1$}{0.45}{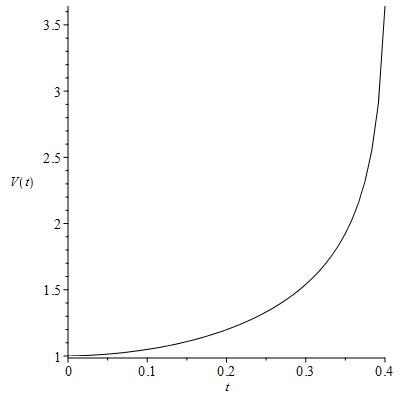}{0.40}{Evolution of $V$ corresponding to the
phase diagram given in Fig. \ref{Fig17} }{0.45}

\myfigures{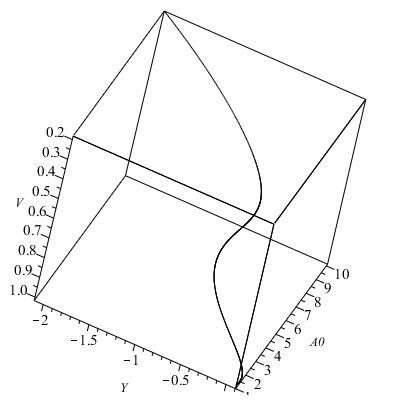}{0.40}{Phase diagram of $[V,\, \dot V,\, A_0^0]$ in
case of $b = 0$, $\lambda_1 = -1$ and $\lambda_2 =
1$}{0.45}{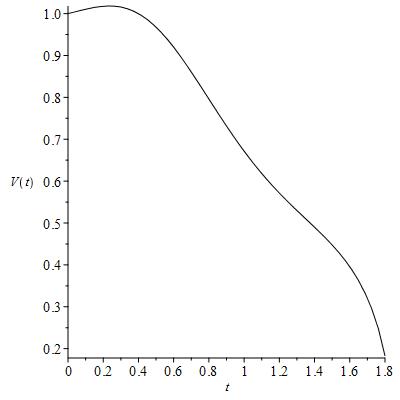}{0.40}{Evolution of $V$ corresponding to the phase
diagram given in Fig. \ref{Fig19} }{0.45}

One of the principal differences between the two cases considered
here is the following. In case of $f^\prime \ne 0$ we have
oscillating mode of expansion for a positive $\lambda_1$, whereas in
case of $f^\prime = 0$ for a positive $\lambda_1$ we have an
accelerated mode of expansion, i.e. the sign of $\lambda_1$ gives
rise two opposite type of evolution.

\section{Conclusion}

Within the scope of Bianchi type-$II$ cosmological models we have
studied the role of nonlinear spinor field in the evolution of the
Universe. In doing so we have considered two cases with $f^\prime
\ne 0$ and $f^\prime = 0$. It was found that the sign of coupling
constant $\lambda_1$ gives rise two different modes of evolution in
the two cases considered.

\vskip 0.1 cm

\noindent {\bf Acknowledgments}\\
This work is supported in part by a joint Romanian-LIT, JINR, Dubna
Research Project, theme no. 05-6-1119-2014/2016.


\begin{thebibliography}{99}




\bibitem{Saha1997GRG} {\it Saha B. and Shikin G.N}  Gen. Relat. Grav.
{\bf 29}, 1099 (1997).


\bibitem{Saha1997JMP} {\it Saha B. and Shikin G.N.} J. Math. Phys. {\bf 38},
5305 (1997).

\bibitem{Saha2001PRD} {\it Saha B.} Phys. Rev. D {\bf 64}, 123501 (2001).


\bibitem{Saha2004aPRD} {\it Saha B.} Phys. Rev. D {\bf 69}, 124006 (2004).

\bibitem{Saha2004bPRD} {\it Saha B. and Boyadjiev T.} Phys. Rev. D {\bf 69}, 124010 (2004).

\bibitem{PopPLB} {\it  Pop{\l}awski N.J.} Phys. Lett. B {\bf 690}, 77 (2010).


\bibitem{PopPRD} {\it  Pop{\l}awski N.J.} Phys. Rev. D {\bf 85},
107502 (2012).

\bibitem{PopGREG} {\it  Pop{\l}awski N.J.} Gen. Releat. Grav. {\bf 44},
1007 (2012).



\bibitem{FabIJTP} {\it Fabbri L.} A Int. J. Theor. Phys. {\bf 52}, 634 (2013).

\bibitem{Saha2006ECAA} {\it Saha B.}  Phys. Part. Nuclei. {\bf 37}, Suppl. 1 S13 (2006).


\bibitem{kremer1} {\it Ribas M.O., Devecchi F.P., Kremer. G.M.}
Phys. Rev. D {\bf 72}, 123502 (2005).



\bibitem{Saha2006PRD} {\it Saha B.}  Phys. Rev. D {\bf 74}, 124030 (2006).


\bibitem{Saha2006GnC} {\it Saha B.} Gravit Cosmology {\bf 12} No 2-3 (46-47)
215 (2006).


\bibitem{Saha2007RRP} {\it Saha B.} Roman. Rep. Phys. {\bf 59},  649 (2007).



\bibitem{Saha2009aECAA} {\it Saha B.} Phys. Part. Nuclei. {\bf 40}, 656 (2009).


\bibitem{ELKO}  {\it  Fabbri L.} Phys. Rev. D {\bf 85}, 047502 (2012).





\bibitem{FabGRG} {\it Fabbri L.} Gen. Relat. Grav. {\bf 43},  1607 (2011).


\bibitem{Krechet} {\it Krechet V.G., Fil'chenkov M.L., and Shikin G.N.}
 Gravit. Cosmology {\bf 14}, No 3(55) 292 (2008).


\bibitem{Saha2010CEJP} {\it Saha B.} Cent. Eur. J. Phys. {\bf 8}, 920 (2010).


\bibitem{Saha2010RRP} {\it Saha B.}  Roman. Rep. Phys. {\bf 62}, 209 (2010).


\bibitem{Saha2011APSS} {\it Saha B.}  Astrophys. Space Sci.
{\bf 331}, 243 (2011).



\bibitem{Saha2012IJTP}{\it Saha B.}  Int.
J. Theor. Phys. {\bf 51}, 1812 (2012).



\bibitem{Saha2015CJP}{\it Saha B.}
Canadian J. Phys. {\bf 93}, 1 (2015).


\bibitem{Saha2015CnJP}{\it Saha B.} Chinese J. Phys. {\bf 53},
110114 (2015).

\bibitem{Saha2016IJTP} {\it Saha B.} Int. J.
Theor. Phys. {\bf 55}, 2259 (2016).


\bibitem{Saha2016EPJP}{\it Saha B.} Eur. Phys. J. Plus. {\bf 131}, 170 (2016).

\bibitem{SahaArXiV2017} {\it  Saha  B.} ArXiV- 1705.07773 [gr-qc]


\bibitem{Suresh} {\it Kumar S. and Akarsu O.} ArXiV 1110.2408 [gr-qc]
(2012)

\bibitem{Singh} {\it Singh J. K. and Sharma N. K.} Int. J. Theor. Phys. {\bf 49}, 2902
(2010)

\bibitem{Reddy} {\it  Reddy D. R. K.,  Patrudu B.M. and Venkateswarlu R.}
Astrophys. Space Sci. {\bf 204}, 155 (1993)


\bibitem{Reddy1} {\it Venkateswarlu R. and Reddy D. R. K.}
Astrophys. Space Sci. {\bf 182}, 97 (1991)


\bibitem{Chauvet} {\it Chauvet P. and  Nunez-Yepez H.N.}
Astrophys. Space Sci. {\bf 178}, 165 (1991)


\bibitem{SahaCEJP2011}{\it  Saha B.} Centr. Euro. J. Phys. {\bf 9}, 939 (2011)

\bibitem{SahaGnC2013}{\it  Saha B.} Gravit. Cosmology {\bf 19}(1), 65
(2013).

\bibitem{SahaAPSS2014} {\it Rikhvitsky V.,  Saha B. and Visinescu M.}
Astrophys.  Space Sci.  {\bf 352}, 255 (2014)

\bibitem{SahaONP2014}{\it Saha B.}
The Open Nuclear $\&$ Particle Physics Journal {\bf 4}, 1 (2011)




\end{thebibliography}
\end{document}